\documentstyle[aas2pp4,12pt]{article}

\def\et{{\it et al.}~}
\def\ms{M_{\odot}}
\def\ers{\rm ergs/sec~}

\begin{document}

\title{SPECTRAL DEPENDENCE OF THE BROAD EMISSION-LINE REGION IN AGN}

\author{Amri Wandel}

\affil{ Racah Institute of Physics, The Hebrew University of Jerusalem 91904, Israel}
\authoremail{amri@vms.huji.ac.il}


\begin{abstract}

We derive a theoretical relation between $R_{BLR}$, the size of the
broad-emission-line region of active galactic nuclei, and the observed soft 
X-ray luminosity and spectrum.
We show that in addition to the well known $R_{BLR}\sim L^{1/2}$ scaling, 
$R_{BLR}$ should depend also on the soft X-ray spectral slope
and derive the expected relation between $R_{BLR}$ and the X-ray luminosity 
and spectral index.
Applying this relation  
to calculate a predicted BLR radius for ten AGN with reverberation data,
 we show that including the dependence on the spectrum  improves
the agreement between the calculated BLR radius and the radius independently 
determined from reverberation mapping.
Similarly we evaluate an expression for the line width,
and show that including the dependence on the spectrum significantly improves
the agreement between the calculated BLR velocity dispersion
and the observed FWHM of the H$\beta$ line.
The theoretical expression for the line width also 
provids a physical explanation to the anti-correlation between the
soft X-ray slope and the emission-line width observed in narrow-line
Seyfert galaxies.

\end{abstract}


\section{INTRODUCTION}

Recent results from reverberation-mapping of the broad emission-line regions 
(BLR) in AGN indicate that the BLR distance from the central radiation source
roughly scales as $r\propto L^{1/2}$ (Koratkar \& Gaskell 1991;Kaspi et.al. 1996; Peterson 1995).
Our model provides a physical 
explanation to this scaling, and predicts an additional testable dependence
on the spectral shape.

Most workers today agree that the line width in AGN is induced by Keplerian 
bulk motions. If this is the case, the BLR distance is directly 
related to the line width, and studying such relations could provide more 
observational input. 
Such an input seems to be found in a new class of
Narrow-Line X-ray Galaxies (NLXG), which are similar to Seyfert 1 galaxies
except for having relatively narrow permitted lines
(FWHM $H_{\beta} < 2000$km/sec) , and often stronger FeII emission 
(e.g. Walter and Fink 1993). 
Many NLXG show steep soft X-ray spectra, $\alpha_x>2$, while ordinary
(broad line) AGN such as Seyfert 1's and quasars usually have 
$\alpha_x\sim 0.5-1.5$. 
Mixed samples (NLXG and ordinary AGN) show a strong correlation between the 
H$\beta$ line width and the soft X-ray spectral index
(Boller, Brandt and Fink 1996; Wang, Brinkmann and Bergeron 1996, hereafter WBB). 
The latter authors found the soft X-ray spectral index 
is correlated also with
the FeII/H$\beta$ ratio, and with the H$\beta$ equivalent width.

We derive simple analytic relations between soft X-ray continuum spectrum
(luminosity and spectral index) and
the BLR size and the width of the broad emission lines, which explain the 
observed correlation between the
 X-ray slope and the line width (section 2).
Applying the model to samples of AGN it gives a good agreement with the BLR 
sizes determined by 
reverberation mapping, and with the observed  FWHM(H$\beta$) - 
cf. Wandel and Boller 1997b ( section 3). 
Finally, in section 4 we suggest explanations of two additional empirical
correlations (related to the H$\beta$ equivalent width and the FeII emission)
and discuss the validity of our basic assumptions.

\section{THE MODEL}

While there is probably a range of physical conditions (ionization
parameter and density etc.) in the broad Hbeta line-emitting gas in AGNs,
the emission of each line is dominated by emission in a relatively narrow
optimum range of conditions (Baldwin et al. 1995).  We will therefore
make the assumption that the conditions relevant for H$\beta$ can be approximated
by a single value.
 We show that within the observational constraints
on the far UV and soft X-ray bands of AGN spectra, the  BLR size and the
line width predicted by the model are not very sensitive to the detailed 
shape of the spectrum in the EUV band.
We then estimate
the BLR size, and  assuming  Keplerian velocity dispersion we relate it
to the line width. The new element in our scheme is the use of the soft 
X-ray spectral index and luminosity to estimate the ionizing spectrum.
 The spectral index enters in the model as follows:
a softer (that is steeper) spectrum has a stronger ionizing power, 
and hence the BLR is formed at a larger distance from the central source,
has a smaller velocity dispersion and produces narrower emission lines.
We parametrize the form of the ionizing EUV continuum in terms of the
 the break between the UV and X-ray bands and the measured 
soft X-ray slope $\alpha_x$. 

\subsection {The BLR radius}

We assume that the spatial extent of the BLR 
may be represented by a characteristic size - e.g. the radius at which 
the emission peaks.  

The physical conditions in the line-emitting gas are largely determined by
the ionization parameter (the ratio of ionizing photon 
density to the electron density, $n_e$, e.g. Netzer, 1990)\qquad
$U = Q_{ion} /{4\pi R^2} c n_e  $
where the ionizing photon flux (number of ionizing photons per unit time) is 
$Q_{ion}= \int_{E_0}^\infty F(E) {dE \over E} $,  
$F(E)$ is the luminosity per unit energy, 
and $E_0=$1 Rydberg=13.6 eV.
Defining the ionizing luminosity,
$L_{ion} = \int_{E_0}^\infty F(E) dE$  
and the mean energy of an ionizing photon,
$\bar E_{ion} \equiv L_{ion}/Q_{ion}$,
the BLR radius (expressed in light-days) may be written as
$$R= \left( {L_{ion}\over 4\pi c\bar E_{ion} U n_e}\right ) ^{1/2}
= 13.4  \left ( {L_{x44}\over U n_{10}  \epsilon_x} \right )^
{1/2}~~{\rm ld} \eqno (1)$$
where $n_{10}=n_e/10^{10}~~ cm^{-3}$, $L_{x44}=L_{x}/10^{44}~$
\ers is the observed X-ray luminosity,
$ \epsilon =\bar E_{ion}/E_0$ is the mean photon energy in Rydbergs,
and $\epsilon_x = \epsilon L_x /L_{ion} = L_x/E_0 Q_{ion}$.
Typical values in the gas emitting the high excitation broad lines are 
$U\sim0.1-1$ and
$n_e \sim 10^{10}-10^{11} cm^{-3}$ (e.g. Rees, Netzer and Ferland, 1989),
so that $Un_{10}\sim 0.1-10$.
For $L_x$ we consider two practical cases:
 broad band luminosity,
(for the ROSAT band) $L_x=\int_{0.1keV}^{2.4keV} F(E) dE$
and a monochromatic luminosity defined  by $L_x=EF(E)$ at E=0.3keV.
The two cases give quite different $R_{BLR}$ vs. $\alpha_x$ curves, 
as shown in fig.1.

\placefigure{fig 1}

\subsection {The ionizing spectrum}

The ionizing flux is dominated by the EUV continuum in the 1-10
Rydberg regime, where most of the ionizing photons are
emitted. Since the the continuum 
in this range cannot be observed directly, we try to estimate it by 
extrapolation from the nearest observable energy bands. 
The far UV spectrum has been observed beyond
the Lyman limit, for about 100 quasars (Zheng et al,
1996), to wavelengths of 600\AA~ , and for a handful
luminous, high redshift quasars to  wavelengths of
350\AA.  The average spectrum has a slope of $\alpha\sim 1$ in the 
1000-2000\AA~
band, and below 900\AA~ it steepens to $\alpha\sim 2$.
In the soft X-ray band the observed spectrum 
is often steeper than this (especially in NLXGs, see Hasinger \et 1993; 
Walter and Fink, 1993, but compare Laor \et 1997).  
This indicated there may be a break or a turnover at
some intermediate energy $E_b$ between 10 and 100eV (cf. Mathews and Ferland
1987).
We assume that the soft X-ray spectrum can be extrapolated to 
lower energies down to some break energy
$E_{b}$, and below the break we take $E^{-2}$.
For the hard X-rays ($E>2keV$) we use the ''universal" $E^{-0.7}$ power law, observed in most AGN
with a high energy cutoff at 100keV
(since the ionizing photon flux is dominated by the continuum 
near 1 Ryd, the model is insensitive to the assumed hard X-ray spectrum).
In summary, we
approximate the ionizing continuum spectrum by a broken power law of the form
$$F(E) \propto \cases{ E^{-2} & $E_{0}<E < E_b$; \cr
 E^{-\alpha_x} & $E_{b}<E < E_h$; \cr
                      E^{-0.7}& $E_h<E<E_{max}$;\cr}  \eqno (2)$$ 
where the break energy $E_{b}\geq E_0$ is a
free parameter,
$E_h$ is the break energy at the hard X-ray band, 
and $E_{max}$  is the high energy cutoff. 
In the calculations below we take $E_b=0.1keV$ and  $E_h=2keV$.

Fig. 1 shows how the calculated BLR size depends on   
the X-ray spectral index, on the shape of the EUV spectrum
(parametrized by $E_b$) and on the normalization
of the X-ray luminosity (broad-band or monochromatic). 
While the dependence on $\alpha_x$
is strong ($R_{BLR}$ increases by a factor of 100 when $\alpha_x$ increases 
from 0 to 4), the dependence on the EUV shape is quite weak;
changing $E_b$ in
 the  relevant range (13.6-100eV) changes $R_{BLR}$ by less than 
30\% (for $\alpha_x<3$).
As expected, we find that $R_{BLR}$ is almost independent of 
the hard X-ray parameters - $E_h$ and $E_{max}$.


\subsection {Line width}
Assuming that the line width is induced by  Keplerian motion in the
  gravitational potential of the central mass, the velocity dispersion 
corresponding to the 
full width at half maximum of the emission lines is given by
%
$v\approx \sqrt {GM/R}$
where M is the mass of the central black hole
and r the distance of the broad line region from the central source.
Eq. (1) gives
$$v\approx 1900~ \left ({Un_{10} \epsilon_x\over L_{x44}}\right )^{1/4}
\left ({M\over 10^7 \ms}\right )^{1/2}~{\rm km/s}. \eqno (3)$$

In order to relate the unknown mass to the observed 
luminosity we assume that 
the central mass approximately scales with the luminosity 
(Dibai 1981; Joly \et 1985; Wandel and Yahil 1985; Wandel and Mushotzky 1986). 
In terms of the Eddington ratio these authors find for large AGN samples
$L/L_{Edd}\approx 0.01-0.1$. 
Within this distribution, bright objects tend to have slightly
larger L/M ratios than faint ones 
(cf. Koratkar and Gaskell 1991) 
roughly $L/M\propto L^{1/4}$ or $M\propto L^{3/4}$.
Combining this with the correlation between the optical and the X-ray
luminosities $L_x\propto L_{opt}^{0.75\pm 0.05}$ 
  (Kriss 1988; Mushotzky and Wandel 1989)
 gives 
$$M\approx 7\times 10^7 L_{x44} 
\left ({ L/L_{Edd} \over 0.01}\right )^{-1} ~\ms .
\eqno (4)$$

and with eqs. (2) and (3)  
the line-width may be related directly to the observed X-ray luminosity 
and spectral index:

$$v(FWHM)\approx 5000 ~\eta \epsilon_x ^{1/4} (\alpha_x )
L_{x44}^{1/4} ~~{\rm km/s}
 \eqno (5)$$
where 
$\eta\equiv (Un_{10})^{1/4}  ( L/L_{Edd} / 0.01 )^{-1/2}$
combines all the unknown parameters; 
for the  $Un_{10}\sim 0.1-10$ and  $L/L_{Edd}\sim 0.01-0.1$ 
stated above we have $\eta \sim 0.2-2$.

\section{COMPARISON WITH THE DATA}
\subsection{BLR radius}
 
We have Compared the BLR size calculated from the model
with the distance from the central source obtained by reverberation mapping  
of the H$\beta$ line,
for a sample of 10 AGN for which reverberation and X-ray data were available 
(fig 2 and table 1). 
As we discuss below, the agreement {\it actually improves} by taking the X-ray 
spectral slope into account.

In the model calculations we have used the spectral index
from the power-law fit to the ROSAT 0.1-2.4 keV band and $E_{b}$=1Ryd, 
(with the exception of NGC 3783, for which we have used
 $\alpha_x=1.5$, 
see Walter and Fink, 1992; note also that NGC 4151 has an extended emission,
Warwick et.al. 1995).
The horizontal error bars represent the combined error in the 
luminosity and in the spectral index.
Allowing for the uncertainty in $Un_{10}\sim 0.1-1$ 
(represented by the dashed lines in fig. 2), the agreement is 
quite good: all the points lie well within these boundaries.
\placefigure{fig 2}
In order to test the significance of our model, we have
 calculated the BLR radii {\it without} taking into account the spectral dependence
(that is, assuming all objects have {\it the same} soft X-ray spectral slope,
which we set to the sample average, $\alpha_x$=1.45).
We find that
the difference between BLR size (calculated
using only the luminosity scaling)  and the reverberation size 
 is significantly larger
when the  spectral dependence is not taken into account
for three objects 
(NGC4151, PG0844 and NGC 3783) shown as gray dots in fig 2.
For the other objects the difference is small in both calculations.

In order to compare our model to the empirical $R\sim L_{opt}^{1/2}$ relation, we
have calculated the BLR radius using  eq. (1)
with $L=L_{opt}$ and $\epsilon=1$.
%
We denote these three models by $ R(L_x,\alpha_x),~~ R(L_x),$ and $R(L_{opt})$.

The statistical significance of each model is tested by the $\chi^2$ statistic,
$$\chi^2= \sum_i {{[\rm log}(R_i({\rm obs})) -{\rm log}(R_i({\rm mod}))]^2
\over \sigma_i^2({\rm obs})+ \sigma_i^2({\rm mod})}.$$
The variance of each object is taken as the sum of the squared standard deviations of the
observed radius (the error in the reverberation determination) and the
calculated radius (the errors in the involved observables - 
$L_x, \alpha_x$ or $ L_{opt}$.
The $\chi^2$ values obtained for the three models are
$ \chi^2(L_x,\alpha_x)=0.88,~~ \chi^2(L_x)= 1.77,$ and $\chi^2(L_{opt})=1.56$.
The corresponding confidence levels $p$ are given by $1-p=\Gamma(\nu/2,\chi^2/2)$ where $\nu$
 is the number of degrees of freedom (here taken as 9).
For the three models we get $1-p$=0.0043, 0.0091 and 0.0077 respectively.

\placetable{tbl-1}

\subsection{Line width-spectral index correlation}

Equation (5) predicts an explicit relation between the velocity dispersion
(associated with the line width), the
X-ray luminosity and the spectral index, namely a surface in the
$\alpha_x - v - L_x$ space. For a fixed value of $L_x$ this gives a curve 
in the $v- \alpha_x$ plane.
Figure 3 shows such curves of FWHM vs.$\alpha_x$  
for several values of the luminosity, $L_{x}=10^{42}$-$10^{45}\ers$
(cf. Wandel and Boller 1997). 
Overplotted are the data points - 
FWHM(H$\beta$) vs. $\alpha_x$ for a sample of AGN (see below).
The  model seems to reproduce the distribution
of the data very well, and in particular it explains the
observed anticorrelation between
the H$\beta$ line width and the soft X-ray spectral index.

\placefigure{fig 3}

\subsection{Predicted vs. observed line width}

In order to test the model prediction over the three dimensional  
$\alpha - v - L_x$ space we compare the observed line width to the
value calculated with eq. (5) using the measured X-ray spectral 
index and luminosity for a sample consisting of 
33 ordinary AGN from Walter and Fink (1993) combined with 32 NLXGs 
(Boller, Brandt and ,Fink 1996). 
As can be seen in fig. 4a, the agreement is very good, and most of
the objects fall well within the uncertainty strip
of log FWHM(obs)=log v(mod)$\pm$0.5.
Fig. 4b shows the same comparison when the spectral information is
not taken into account (but rather its sample average, $\alpha_x=1.74$): 
the correlation is significantly weeker.
The correlation coefficient between the observed and calculated line widths is
$r= 0.533$ and 
0.316 respectively.
The t-test  statistic for zero regression 

$t=r \sqrt{(N-2)}/\sqrt{(1-r^2)}$ gives for the two cases 
$t=3.60$ and 2.40 respectively, for which the confidence levels for zero 
correlation
are $1-p=0.0004$ and 0.01 respectively, and the dependent confidence level (which 
measures the significance of the dependence on spectral index, after taking
out the dependence on the luminosity) is $p= 1-0.0004/0.01=0.96$.

\placefigure{fig 4ab}

%
\subsection{Equivalent width}
One may try to understand also the other strong correlations found in the data
in terms of our model.
The H$\beta$ equivalent width is anti-correlated with the FWHM (Gaskell 1985, WBB) and with the
soft X-ray spectral 
index; AGN with a steep soft X-ray spectrum have a lower EW(H$\beta$) than flat-
spectrum AGN. To see how our model can explain this, we recall that 
the equivalent width measures the fraction of
the continuum flux reprocessed and emitted in the line.
We have shown (fig. 1) that the BLR distance from the central source
increases with the spectral index. 
If the emission-line clouds are not pressure confined,
it is reasonable that their sizes do not change significantly with the BLR radius, 
hence if clouds are conserved, the solid angle covered by the clouds decreases with radius, 
which implies that the equivalent width decreases 
with increasing (steepening) spectral index, as observed.
%
\acknowledgments
The author thanks Th. Boller, A. Laor, 
D. Maoz and H. Netzer for useful discussions, M.I. for help with the 
statistical analyses and M. Gaskell for the careful refereeing.

\vskip 2 cm

\begin{deluxetable}{cllllllll}
\footnotesize
\tablecaption{X-ray and Optical data, reverberation and model BLR radius. 
\label{tbl-1}}
\tablewidth{0pt}
\tablehead{
\colhead{Object} &\colhead{Name} & \colhead{Log $L_{X}$}   &\colhead{Log $L_{opt}$} & 
\colhead{$\alpha_x$}   &  \colhead{$R_{rev}(H\beta )$\tablenotemark{a}} &
 \colhead{$R(L_x,\alpha_x)$\tablenotemark{b}} &
 \colhead{$R(L_x)$\tablenotemark{b}} &
 \colhead{$R(L_{opt})$\tablenotemark{b}} 
} 
\startdata

1 &NGC 3783& 43.5$\pm$0.16& 43.5$\pm$0.2& 1.50$\pm$0.40& 0.92$\pm$0.08& 0.91$\pm$0.29& 0.88$\pm$0.08& 0.90$\pm$0.1\nl
2 &NGC 5548& 44.0$\pm$0.06& 43.8$\pm$0.2& 1.21$\pm$0.15& 1.23$\pm$0.10& 1.00$\pm$0.11& 1.13$\pm$0.03& 1.05$\pm$0.1\nl
3 &MKN  279& 44.26$\pm$0.07& 44.2$\pm$0.1& 1.15$\pm$0.17& 1.00$\pm$0.10& 1.10$\pm$0.13& 1.26$\pm$0.035& 1.24$\pm$0.05\nl
4 &AK   120& 44.86$\pm$0.08& 44.15$\pm$0.25& 1.63$\pm$0.29& 1.46$\pm$0.15& 1.66$\pm$0.19& 1.56$\pm$0.04& 1.20$\pm$0.12\nl 

5 &MKN  590& 44.2$\pm$0.08& 44.3$\pm$0.2& 1.01$\pm$0.24& 1.27$\pm$0.04& 1.00$\pm$0.17& 1.23$\pm$0.04& 1.30$\pm$0.1\nl
6 &PG  0804& 45.2$\pm$0.1& 45.3$\pm$0.1& 1.53$\pm$0.41& 1.92$\pm$0.23& 1.78$\pm$0.26& 1.73$\pm$0.05& 1.80$\pm$0.06\nl
7 &NGC 4541& 43.1$\pm$0.1& 43.2$\pm$0.2& 1.80$\pm$0.1& 0.85$\pm$0.25& 0.87$\pm$0.10& 0.68$\pm$0.05& 0.74$\pm$0.1\nl

8 &PG  0953& 45.16$\pm$0.05& 45.7$\pm$0.2& 1.31$\pm$0.28& 1.95$\pm$0.20& 1.64$\pm$0.17& 1.71$\pm$0.025& 1.98$\pm$0.1\nl
9 &PG  0844& 43.7$\pm$0.08& 45.4$\pm$0.3& 1.82$\pm$0.95& 1.57$\pm$0.27& 1.17$\pm$0.50& 0.98$\pm$0.04& 1.83$\pm$0.15\nl
10&NGC 3516& 42.7$\pm$0.03& 43.5$\pm$0.1& 1.49$\pm$0.42& 0.59$\pm$0.12& 0.50$\pm$0.23& 0.48$\pm$0.015& 0.88$\pm$0.05\nl

\enddata

\tablenotetext{a}{Log of BLR radius from reverberation mapping in light days}
\tablenotetext{b}{Log of BLR radius calculated from the model in light days (see text)}

\end{deluxetable}

{\bf Figure Captions}

\figcaption[fig 1.ps]{ The calculated BLR radius vs. the soft X-ray spectral index $\alpha$. 
Shown are two assumed EUV spectra:
(a) $E_b=1$Ryd, so that $\alpha_{EUV}=\alpha_x$ 
(solid curves), and (b) $E_b=100eV$, 
$\alpha_{EUV}=2$ (dashed curves).  
For each case we show two luminosity normalizations:  monochromatic -
$L_x(0.3keV)=10^{44}\ers$ (the {\it upper} two curves, indicated by 0.3keV) and 
the ROSAT band normalization (see text) - $L_x(0.1-2.4keV)=10^{44}\ers$ 
(the {\it lower} two  curves, indicated by 0.1-2.4keV).
\label{fig 1}}

\figcaption[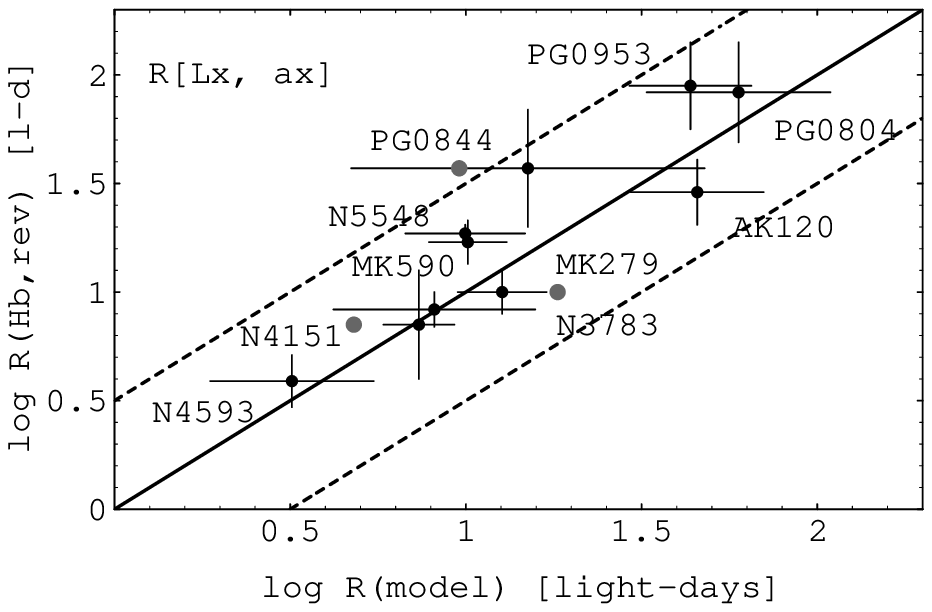] {The BLR size determined from reverberation mapping of the
H$\beta$ line vs. the radius calculated from the model (eq. (1)) using
the observed X-ray luminosity and spectral index.
The reverberation radii are from Kaspi et.al. (1996), and the X-ray data from
Walter and Fink (1993) except NGC 4151 (Warwick \et 1995)
and PG 0026 and PG 0844 from WBB.
The grey circles represent the  radius calculated from the model without
taking the spectral dependence into account (see text).
The solid line represents $R_{rev}=R_{model}$ for $Un_{10}=1$
while the dashed parallel lines above and below it 
correspond to $Un_{10}=0.1$ and $10$ respectively.
\label{fig 2}}

\figcaption[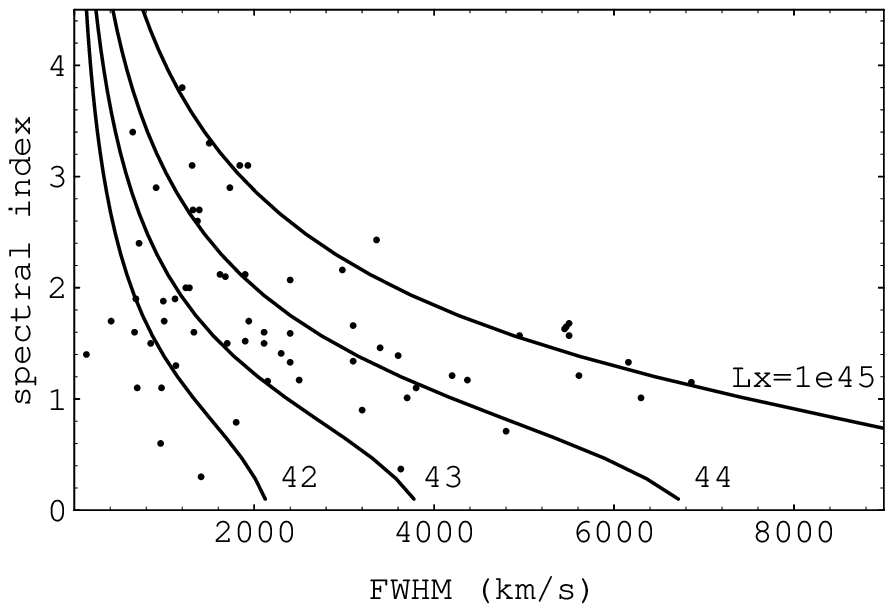]{ Theoretical curves of the X-ray spectral index vs. v(BLR) for 
fixed values of $L_x=10^{42}\ers$ (lower curve) to $L_x=10^{45}\ers$ (upper curve)
superimposed on the data for the 
sample of NLXGs and normal Seyferts. \label{fig 3}}

\figcaption[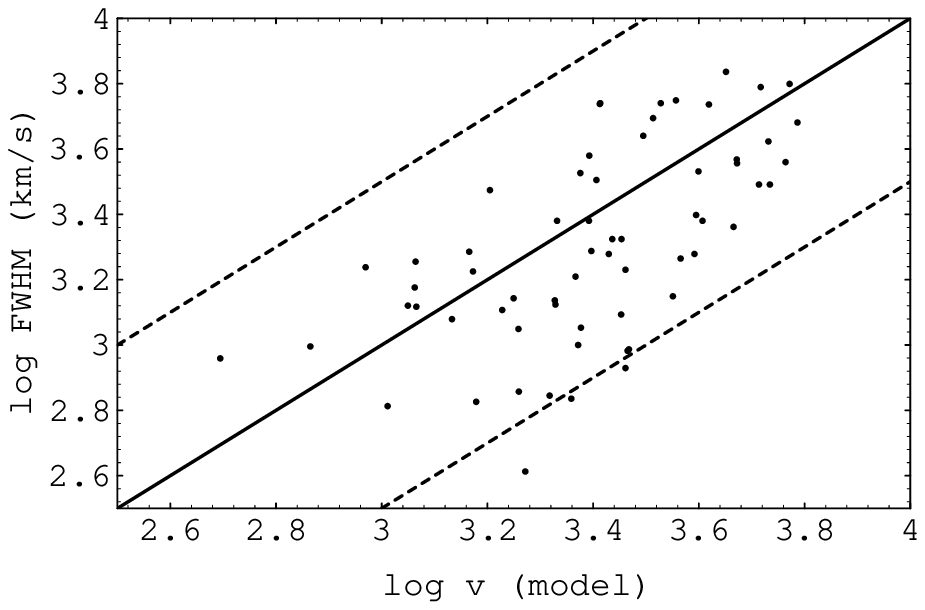,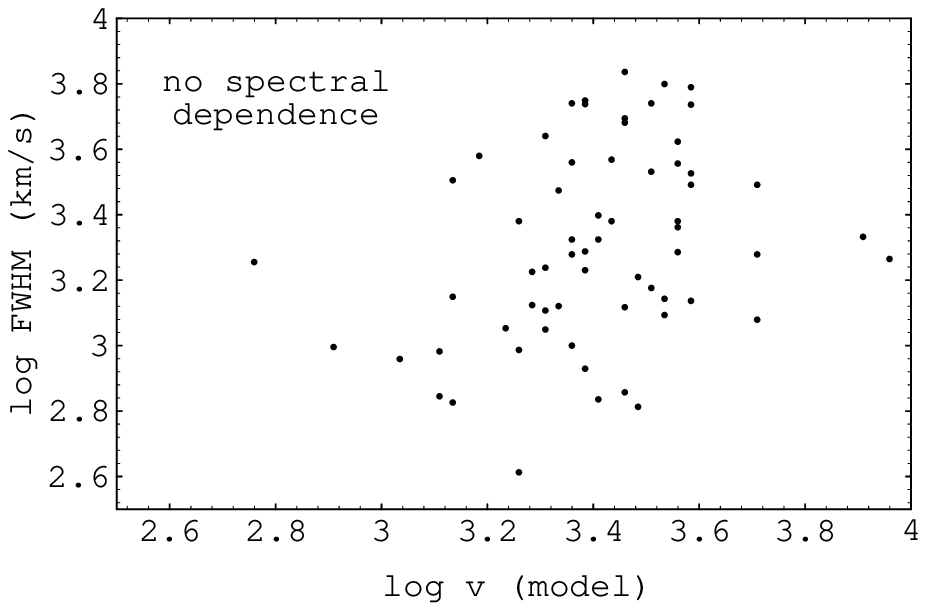]
{Observed FWHM(H$\beta$) vs. $v(L_x,\alpha_x)$ (the velocity dispersion 
calculated with the observed X-ray spectral index and luminosity, eq. (5))
 - figure 4a, and
 FWHM(H$\beta$) vs. $v(L_x)$ (calculated with only the observed X-ray 
luminosity- figure 4b.
The diagonal line represents the equality ($v=FWHM$) for $\eta =0.6$, 
while the dashed parallel lines above and below the diagonal are for
$\eta=0.2$ and 2 respectively.\label{fig 4ab}}

\end{document}